\def\HI{\hbox{H {\sc i}}}

\documentclass[11pt,twoside]{article}
\usepackage{newpasp}
\usepackage{epsf}

\index{Buote, D. A.}

\begin{document}

\title{Warm Absorbing Gas in Cooling Flows}
\author{David A. Buote}
\affil{UCO/Lick Observatory, University of California,
Santa Cruz, CA 95064}


\begin{abstract}
We summarize the discovery of oxygen absorption and warm ($10^5-10^6$
K) gas in cooling flows.  Special attention is given to new results
for M87 for which we find the strongest evidence to date for ionized
oxygen absorption in these systems. We briefly discuss implications
for observations of cooling flows with {\sl Chandra} and {\sl XMM}.
\end{abstract}




\section{Introduction}

The inhomogeneous cooling flow scenario is often invoked to interpret
the X-ray observations of massive elliptical galaxies, groups, and
clusters.  The key prediction of this scenario is the existence of
large quantities of gas that have cooled out of the hot phase and
dropped out of the flow. The only evidence for large amounts of mass
drop-out arises from the excess soft X-ray absorption from cold gas
found for many cooling flows especially from spectral analysis of {\sl
Einstein} and {\sl ASCA} data. This interpretation is highly
controversial because for systems with low Galactic columns no excess
absorption from cold gas is ever found with the {\sl ROSAT} PSPC which
should be more sensitive because of its softer bandpass, 0.1-2.4
keV. Furthermore, the large intrinsic columns of cold H indicated by
the {\sl Einstein} and {\sl ASCA} data are in embarrassing
disagreement with \HI\, and CO observations. We have re-examined the
{\sl ROSAT} PSPC data of cooling flows to search for evidence of
intrinsic soft X-ray absorption and in particular have allowed for the
possibility that the absorber is not cold.

\section{M87}

\begin{figure}[t!]
\vbox {
  \begin{minipage}[l]{1.0\textwidth}
   \hbox{
       \begin{minipage}[l]{0.51\textwidth}
       {\centering \leavevmode \epsfxsize=\textwidth
        \epsfbox{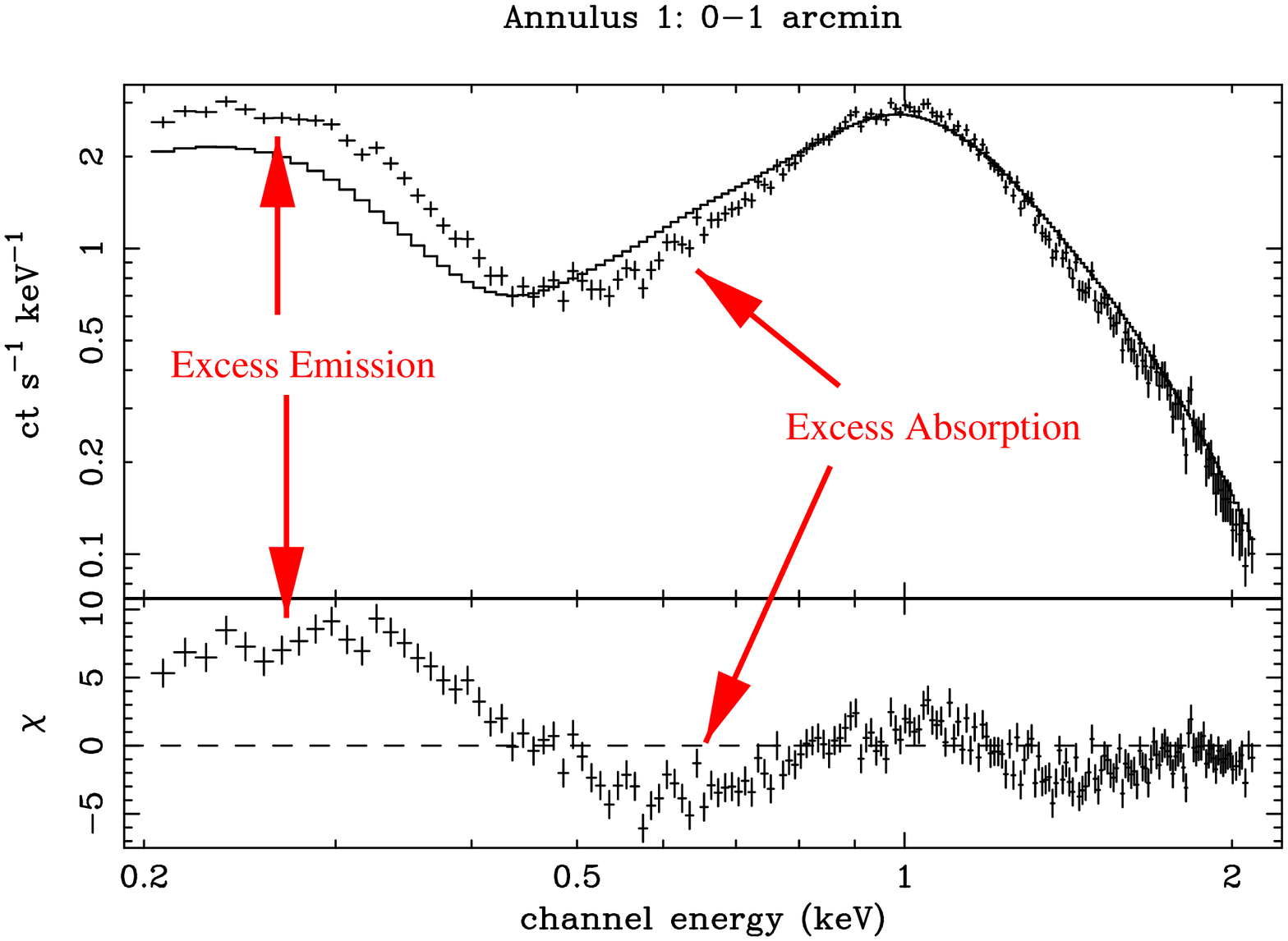}}
       \end{minipage} \  \hfill \
       \begin{minipage}[r]{0.46\textwidth}
       {\centering \leavevmode \epsfxsize=\textwidth
        \epsfbox{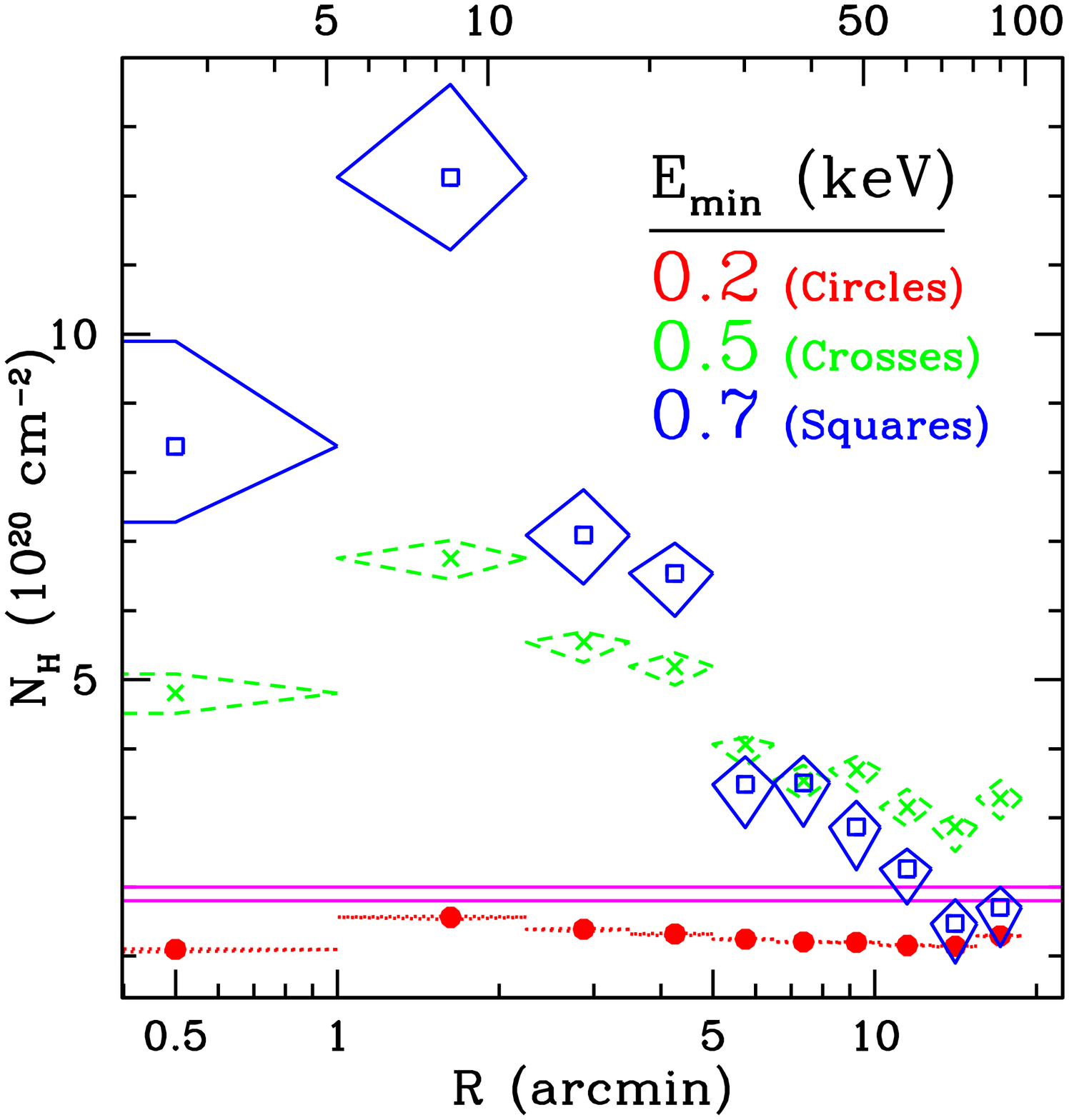}}
       \end{minipage} \  \hfill \
  }
  \end{minipage} \  \hfill \
  \begin{minipage}[l]{1.0\textwidth}
\caption{\label{fig} \small (Left) {\sl ROSAT} PSPC spectrum of the
the central arcminute of M87. The fitted model has hydrogen column
density fixed to Galactic and Fe abundance fixed to $0.78Z_{\odot}$.
(Right) Radial profiles of hydrogen column density of the standard
foreground absorber for different values of the minimum energy of the
bandpass for single-phase models. Only data with energies $E>E_{\rm
min}$ are included in the fits. The solid lines (magenta) indicate the
range of Galactic column density. Both figures are taken from Buote
(2000c).}
\end{minipage} }
\vspace*{-0.4cm}
\end{figure}            

M87 is arguably the best target for study of its hot plasma with {\sl
ROSAT} because it is the brightest nearby galaxy, group, or cluster
that also possesses an ambient gas temperature ($\sim 2$ keV) that
lies within the bandpass of the PSPC. From analysis of spatially
resolved, deprojected {\sl ROSAT} PSPC spectra we find the strongest
evidence to date for intrinsic oxygen absorption and multiphase gas in
the hot ISM/IGM/ICM of a galaxy, group, or cluster. When attempting to
describe the 0.2-2.2 keV {\sl ROSAT} emission of M87 by a single-phase
hot plasma modified by standard Galactic absorption the best-fitting
model displays striking residuals in the spectrum (Figure \ref{fig}):
(1) excess emission above the model for 0.2-0.4 keV and (2) excess
absorption below the model for 0.5-0.8 keV. These features are
apparent out to the largest radii investigated ($\sim 100$ kpc) and
cannot be attributed to errors in the calibration or the background
subtraction.

The principal result is that the 0.5-0.8 keV absorption is consistent
with that of a collisionally ionized plasma with a temperature of
$10^{5-6}$ K, where the lack of evidence of absorption below 0.5 keV
strongly excludes the possibility of absorption from cold material as
has been assumed in essentially all previous studies of absorbing
material in cooling flows. In fact, the excess {\it emission} observed
between 0.2-0.4 keV, which is also manifested as sub-Galactic column
densities in models with standard (cold) absorbers, has a temperature
that is consistent with the emission from gas responsible for the
0.5-0.8 keV absorption and could not be explained if dust were
responsible for both the absorption and soft emission. 

Only multiphase models can provide good fits over the entire PSPC
bandpass while also yielding temperatures and Fe abundances that are
consistent with results from {\sl ASCA} and {\sl SAX} at large
radii. Both cooling flow and two-phase models indicate that the
fraction of warm gas with respect to the total emission measure
differs qualitatively for radii interior and exterior to $\sim
5\arcmin$ ($\sim 26$ kpc). For $r>5\arcmin$ the data are consistent
with a constant (or slowly varying) fraction of warm gas as a function
of radius. But for $r<5\arcmin$ the warm gas fraction varies from
$\sim 20\%$ within the central bin $(1\arcmin)$ to essentially zero
within the next few bins (out to $5\arcmin$). 

This behavior is to be contrasted with the absorption optical depth
profiles which are approximately constant with radius.  It is puzzling
why the oxygen absorption optical depth does not dip between $R\sim
1\arcmin - 5\arcmin$ as would be expected if the absorption and excess
soft emission arise from the same material. Perhaps this is a result
of the simplifications we have employed (e.g., only assume a single
absorption edge), and with rigorous consideration of the radiative
transfer and the absorption from several ionization states of oxygen
(and carbon and nitrogen) a self-consistent description of the
multiphase medium will be obtained.

The oxygen absorption and soft emission from warm ($10^{5-6}$ K) gas
in M87 we have detected using the {\sl ROSAT} PSPC (0.2-2.2 keV) is
also able to satisfy the detections of excess emission with data at
lower energies (0.07-0.25 keV) from {\sl EUVE} (Lieu et al. 1996) and
the detection of excess absorption at higher energies (0.5-10 keV)
from {\sl Einstein} and {\sl ASCA}. Previous studies using only {\sl
EUVE} data could not decide between a non-thermal and thermal origin
for the excess soft emission.  Similarly, the previous detections of
absorption with {\sl Einstein} and {\sl ASCA} could not constrain the
temperature of the absorber and always assumed a cold absorber with
solar abundances. Even previous studies with {\sl ROSAT} that
neglected data below $\sim 0.5$ keV also could not constrain the
temperature of the absorber (Sanders et al. 2000) because of the
strong dependence of the inferred absorption on the lower energy limit
of the bandpass (Figure \ref{fig}).

Hence, the {\sl ROSAT} detection of intrinsic absorption that is
localized in energy (0.5-0.8 keV) is the key piece of evidence for
establishing the presence of warm ($10^{5-6}$ K) gas distributed
throughout (at least) the central 100 kpc. This evidence for a
multiphase ISM in M87 essentially confirms the original detection
within the central $\sim 2\arcmin$ using the {\sl Einstein}
FPCS. However, instead of intrinsic oxygen absorption Canizares et
al. (1982) inferred a super-solar O/Fe ratio which does not agree with
subsequent analyses of M87 using other instruments. The anomalous O/Fe
ratio is probably attributed to either calibration error in the FPCS
or to an underestimate of the continuum due to the absorption edges.

The total mass of the warm gas implied by the oxygen absorption is
consistent with the amount of matter deposited by an inhomogeneous
cooling flow. On the other hand, the mass deposition profile and the
profile of warm emission fraction of the two-phase models indicate
that the emission of the warm component is suppressed over $r\sim
1\arcmin - 5\arcmin$ where the radio emission from the AGN jet clearly
distorts the X-ray isophotes (Owen et al. 2000). This coincidence
suggests that the AGN has influenced the hot ISM in these central
regions and may have suppressed the cooling emission of the warm
component (see Binney 1999). Within the central arcminute the gas has
apparently readjusted and is cooling while at large radii,
$r>5\arcmin$, the cooling flow was not disturbed significantly by the
AGN. A hybrid model of a standard cooling flow with AGN feedback seems
promising for M87.

\section{Other Cooling Flows}

The highly significant detection of intrinsic oxygen absorption in the
{\sl ROSAT} PSPC data of M87 confirms the picture of a multiphase
warm+hot medium in cooling flows deduced from the lower S/N PSPC data
of galaxies, groups, and the cluster A1795 in Buote (2000a,b).  Unlike
M87 the absorption and soft emission features associated with the
warm gas are not obvious upon visual examination of the lower S/N PSPC
spectra of these systems (e.g., see Figure 5 of Buote 2000b), but five
galaxies and groups (NGC 507, 1399, 4472, 4649, and 5044) and the
cluster A1795 clearly show the same type of sensitivity of $N_{\rm H}$
to the lower energy limit of the bandpass (Figure 1).  Most of these
lower S/N systems also have sub-Galactic values of $N_{\rm H}$ when
fitted to the entire bandpass consistent with excess soft emission
like M87. Hence, warm absorbing gas appears to be common in cooling
flows ranging from galactic to cluster scales.

\section{Observing Oxygen Absorption with Chandra and XMM}

With the launches of {\sl Chandra} and {\sl XMM} it is an advantageous
time to study oxygen absorption and warm ionized gas in cooling
flows. The {\sl XMM} (EPIC) and {\sl Chandra} (ACIS-S) CCDs both have
substantially better energy resolution than the PSPC and, unlike {\sl
ASCA}, both also have bandpasses that extend down to 0.1-0.2 keV
important for measuring the emission from the warm gas.

At the moderate spectral resolution of the {\sl Chandra} and {\sl XMM}
CCDs the absorption signature of the warm gas is expected to be a
relatively broad trough over energies $\sim 0.5- 0.8$ keV: {\it thus
future Chandra and XMM observations will not see a single sharp
feature.} This broad feature is the result of the cumulative
absorption of edges of primarily oxygen but with significant
contributions from C and N (see Krolik \& Kallman 1984 for a model of
a warm absorber in coronal equilibrium).  Perhaps the simplest means
to initially quantify the warm absorption would be to show the
sensitivity of $N_{\rm H}$ to $E_{\rm min}$ for a standard absorber
model (Figure 1).

The grating spectrometers of {\sl XMM} and {\sl Chandra} have even
better energy resolution (but smaller effective area) and, in
principle, might detect individual absorption edges. These gratings
are strictly valid only for point sources and thus will only be able
to obtain spectra for the very central regions of cooling flows
($<\sim 10\arcsec$). Since it is very likely that the warm absorbing
gas emits over a range of temperatures and does not obey strict
coronal equilibrium, the high resolution spectra offered by the
gratings may detect departures from simple isothermal coronal
absorption models and thus provide vital insights into the
thermodynamic state of the warm gas.





\begin{references}
{\small

\reference Binney, J. J. 1999, in The Radio Galaxy M87,
ed. H.-J. R\"{o}ser \& K. Meisenheimer, (Springer-Verlag), 136
\reference Buote, D. A. 2000a, \apj, 532, L113
\reference Buote, D. A. 2000b, \apj, in press (astro-ph/0001330)
\reference Buote, D. A. 2000c, \apj, in press (astro-ph/0008408)
\reference Canizares, C. R., Clark, G. W., Jernigan, J. G., Markert, T. H. 1982, \apj, 262, 33
\reference Krolik, J. H., \& Kallman, T. R., 1984, \apj, 286, 366
\reference Lieu, R., Mittaz, J. P. D., Bowyer,
S., Lockman, F. J., Hwang, C.-Y., \& Schmitt, J. H. M., 1996, \apjl,
458, L5
\reference Owen, F. N., Eilek, J. A., \& Kassim, N. E. 2000, \apj, in
press (astro-ph/0006150)     
\reference Sanders, J. S., Fabian, A. C., \& Allen, S. W. 2000, \mnras, in press (astro-ph/0006394)         



}
\end{references}
\end{document}